\documentclass[runningheads]{llncs}
\usepackage{graphicx}
\usepackage{orcidlink}
\usepackage[numbers,sectionbib]{natbib}
\hypersetup{hidelinks}

\usepackage{url}
\usepackage{subcaption}
\usepackage{soul}
\usepackage{booktabs}
\begin{document}
\title{Supporting architecture evaluation for ATAM scenarios with LLMs}

%\titlerunning{Helping novice architects using an LLM-based assistant}
%\titlerunning{Supporting architecture evaluation for ATAM scenarios with LLMs}
%\author{Anonymous authors}

\author{Rafael Capilla\inst{1}\orcidlink{0000-0002-6943-1285} \and J. Andrés Díaz-Pace\inst{2}\orcidlink{0000-0002-1765-7872} \and
Yamid Ramírez\inst{1}\orcidlink{} \and
Jennifer Pérez\inst{3}\orcidlink{} \and Vanessa Rodríguez-Horcajo\inst{3}\orcidlink{} }
\institute{Rey Juan Carlos University, Madrid, Spain\\
\email{rafael.capilla@urjc.es,ye.ramirez.2024@alumnos.urjc.es}
\and
ISISTAN, CONICET/UNICEN, Tandil, Buenos Aires, Argentina\\ 
\email{andres.diazpace@isistan.unicen.edu.ar}\\
\and
Universidad Politécnica de Madrid, Spain\\
\email{jenifer.perez@upm.es,vanessa.rodriguez.horcajo@upm.es}}

\maketitle              % typeset the header of the contribution
\begin{abstract}
Architecture evaluation methods have long been used to evaluate software designs. Several evaluation methods have been proposed and used to analyze tradeoffs between different quality attributes. Having competing qualities leads to conflicts for selecting which quality-attribute scenarios are the most suitable ones that an architecture should tackle and for prioritizing the scenarios required by the stakeholders. In this context, architecture evaluation is carried out manually, often involving long brainstorming sessions to decide which are the most adequate quality scenarios. To reduce this effort and make the assessment and selection of scenarios more efficient, we suggest the usage of LLMs to partially automate evaluation activities. As a first step to validate this hypothesis, this work studies \texttt{MS Copilot} as an LLM tool to analyze quality scenarios suggested by students in a software architecture course and compares the students' results with the assessment provided by the LLM. Our initial study reveals that the LLM produces in most cases better and more accurate results regarding the risks, sensitivity points and tradeoff analysis of the quality scenarios. Overall, the use of generative AI has the potential to partially automate and support the architecture evaluation tasks, improving the human decision-making process.% accelerating the time spent on architecture evaluation tasks.  

\end{abstract}
\section{Introduction}\vspace{-0.15cm}
For more than two decades, software architects have used different architecture evaluation methods (e.g. ALMA, ATAM ARID, SAAM) \cite{clements2001} to assess quality attributes and determine which and how certain quality-attribute properties 
should be explicitly considered in the architecture before implementing the system. 
Addressing quality attributes 
requires one to identify risks and non-risks of quality scenarios and also the impact of sensitivity points on the architecture \cite{cervantes2024}. Furthermore, tradeoffs between competing qualities may complicate the selection and prioritization of scenarios. This activity often leads to long brainstorming sessions between stakeholders and architects. 

In order to partially automate the evaluation quality scenarios and double-check if the selected scenarios in an architecture evaluation are adequate, 
we propose the usage of generative AI techniques like Large Language Models (LLMs). In addition to making brainstorming sessions more effective, our goal is to assist architects
in their evaluation tasks by suggesting
the most suitable scenarios for improving an architecture based on their pros and cons, alerting architects about risks, and also possible quality-attribute tradeoffs. 

In this initial research work, we investigate how an LLM like \texttt{MS Copilot} can support quality-driven architecture evaluations. To this end, we adopted a Retrieval Augmented Generation (RAG) strategy \cite{lewis2020retrieval} to feed an LLM using an architecture assignment from an undergraduate course of computer science students that applied the Architecture Tradeoff Analysis Method (ATAM) \cite{clements2001} and then we use prompts to instruct the LLM to assess %certain qualities of 
the architecture in a stepwise manner.

The remainder of this paper is as follows. Section 2 describes the background to understand the relevant concepts. Section 3 outlines the study design and research goals. Section 4 describes the initial evaluation performed, and Section 5 discusses the conclusions and future work.

\vspace{-0.45cm}
\section{Background}\label{sec:background}
\vspace{-0.25cm}
ATAM \cite{ATAM2000} is a manual method that aims to assess the consequences and impact of architectural decisions motivated by quality-attribute requirements. An ATAM team is responsible for analyzing the quality needs of stakeholders and producing the so-called \textit{utility tree}, which is formed by a set of quality scenarios addressing specific quality concerns. % (i.e. quality attributes). 
These qualities can impact different system's areas and on the software architecture as well. In addition, each quality scenario is characterized by \textit{stimuli} to which an architecture (or system) must respond and a way to measure the quality-attribute response triggered by the stimuli. The ATAM team must suggest, evaluate, and select the most suitable scenarios by evaluating possible risks, sensitivity points and tradeoffs between qualities. 

In particular, the tradeoffs for each scenario reflect the tensions between two or more qualities, thus a criterion must be used to prioritize and select between several scenarios during the brainstorming sessions. An initial experience analyzing tradeoff and sensitivity points using the Analytical Hierarchical Process (AHP) method is described in \cite{Ibrahim2009}, where the authors rely on pairwise comparisons to support decision-making of commercial off the self (COTS) components. The use of AHP to evaluate the quality attributes of the system architecture according to the ISO 25010:2011 quality model\footnote{Now replaced by the ISO 25002:2024 \url{https://www.iso.org/standard/78175.html}} is also highlighted in \cite{Darwish2017}. Only a few approaches, such as those mentioned in \cite{Lytra2020}, support some kind of automation with respect to architecture quality tradeoffs in multi-attribute decision-making.

Using generative AI % (i.e. Generative Artificial Intelligence) 
and LLMs to make decisions is a timely alternative to speed up architects' decisions and refine system architectures \cite{Ozkaya2023a}. Not far ago, LLMs have been trained and fine-tuned using datasets \cite{brown2020language} in order to respond to natural language questions. 
Therefore, the use of LLM-based tools with adequate prompts \cite{White2023} %to improve decision-making according with adequate prompts \cite{White2023} 
is an important challenge, so that architects do not overlook relevant knowledge when evaluating quality scenarios. %Not far ago, LLMs have been trained and fine-tuned using a dataset \cite{brown2020language} in order to respond to natural language questions. 

\vspace{-0.3cm}
\section{Study Design}\label{sec:approach}
\vspace{-0.2cm}
This work evaluates whether generative AI can enhance human decision-making and reflective practices \cite{Razavian2020} based on prior results of novice architects. We analyzed the results of $9$ student projects from a software architecture course at the university level. In this project, a team of students acting as an ATAM team plus another one acting as a customer captured, evaluated, and selected quality scenarios to incorporate certain quality properties in an existing software architecture. Note that although the case study to produce the architecture in a previous assignment was the same, the final architecture produced by each team might exhibit differences. In addition, the quality needs stated by each student acting as the ``customer'' in the ATAM process can be different in each of the assignments. 

The steps that the students performed to apply ATAM were as follows: (i) they used a software architecture designed in a previous assignment as input, and one student acting as the customer indicated a set of quality needs to improve the current architecture, %quality of the architecture, 
(ii) the quality needs were discussed and refined with $5$ students acting as the ATAM team, (iii) the students followed the ATAM method to create a utility tree describing quality scenarios for each of the quality attributes derived from the system quality goals and affecting different parts of the existing architecture, (iv) each scenario was evaluated to identify possible risks, sensitivity points, and tradeoffs between quality attributes affecting the scenario and (v) the final scenarios selected by students were %modeled
engineered in the architecture. All the information from the students' assignments was gathered in PDF documents. We provided these documents as input to \texttt{MS Copilot} \footnote{\url{https://copilot.microsoft.com/}} and used a RAG strategy\footnote{RAG integrates external domain-specific sources
into the LLM to improve the accuracy and relevance of the response} \citep{lewis2020retrieval} to evaluate the degree to which the scenarios chosen by the students were adequate. An example of the results generated by the LLM using the sequence of four prompts is given here\footnote{\url{https://anonymous.4open.science/r/archevaluation-llms-1FDE/prompting-example.md}}.% the right ones.  

\begin{figure}[ht]
\vspace{-0.2cm}
\centering
\includegraphics[width=0.8\textwidth]{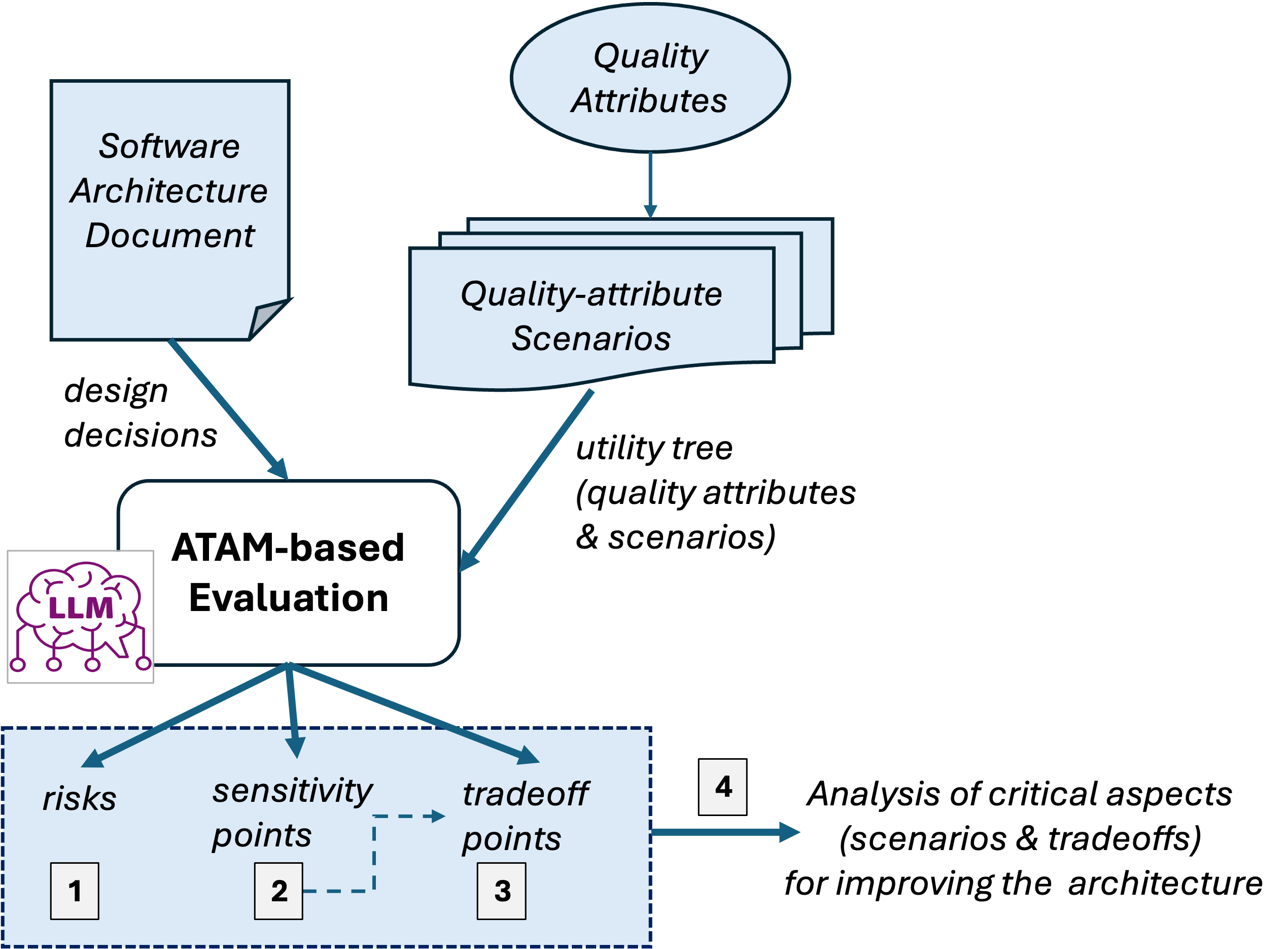}
\caption{Inputs and outputs of the ATAM process, including the steps where the LLM prompts were used ($1$, $2$, $3$, $4$)} 
\label{fig:llm-atam}\vspace{-0.65cm}
\end{figure}

The co-authors tested the validity and accuracy of the scenarios using four different prompts for \texttt{MS Copilot}. The steps of the ATAM process to evaluate and select the quality scenarios are depicted in Figure \ref{fig:llm-atam}, enriched with the 4 prompts used for \texttt{MS Copilot}.% to check if the scenarios chosen by the students where the right ones. 
In particular, we prompted the LLM to identify possible risks, sensitivity points and perform a tradeoff analysis. Afterwards, we asked \texttt{MS Copilot} to perform a selection of scenarios based on the previous information. From the $9$ student groups evaluated, we discarded two groups %(i.e. G5 and G9) 
because the quality of the scenarios was rather low and they provided several unclear scenarios.  % support the identification of risks, sensitivity points and tradeoff analysis. 

\vspace{-0.3cm}\section{Preliminary Evaluation}\label{sec:evaluation}\vspace{-0.2cm}
In this section we describe the results of our LLM-approach to simulate the ATAM process. The scenarios selected by the students were based on their analysis of risks for the scenarios, identification of sensitivity points in the architecture that could affect the scenarios, and a tradeoff analysis between quality attributes. % in some of the scenarios. 
Thus, we used four prompts to query \texttt{MS Copilot} about these aspects to validate the scenarios, in addition to asking the LLM to suggest scenarios. % suggested by \texttt{MS Copilot }in Table \ref{tab:atam}.
%In this section we describe the results of the four prompts used to simulate the ATAM process to evaluate quality scenarios. 
We provide a reproducibility kit that contains the input artifacts and prompts. %, as well as the outputs of \texttt{MS Copilot}, is provided at \footnote{\url{https://anonymous.4open.science/r/archevaluation-llms-1FDE/}}.

\textbf{Prompts \#1 and \#2. Identification of risks and sensitivity points:} In this step we asked \texttt{MS Copilot} to infer risks related to all the scenarios described by the students. % including both the selected and the non-selected scenarios. 
In the context of ATAM, we refer to a risk as a problematic architectural decision that can potentially lead to negative consequences if not addressed properly.
A sensitivity point, in turn, is a design decision that affects the achievement of a particular quality-attribute scenario. %, in the sense that the decision might cause variations in the response of the scenario

The results are shown in Table \ref{tab:risk-sensitivity}, which distinguishes the risks and sensitivity points found by the students and by \texttt{MS Copilot}. As we can observe, \texttt{MS Copilot} was extremely useful in detecting additional risks and sensitivity points (i.e., items not identified by the students), which are relevant for the selection of scenarios. %Therefore, in Table \ref{tab:risk-sensitivity} we provide the risks and sensitivity points found by the students and by the Copilot. 
In several groups, there was a good intersection (e.g. $G6$) while in other groups the LLM detected more risks and sensitivity points (e.g., $G1$, $G3$, $G8$). In those cases where \textit{MS Copilot} suggested additional risks, we list such scenarios in the last column of Table \ref{tab:risk-sensitivity} in order to indicate that in some cases certain scenarios should not have been chosen by the students. However, this decision depends on factors such as the risk severity, probability of occurrence, or business importance of the scenario, because not all risks are equally critical. If a risk identified for a chosen scenario is not critical, we assume that selecting such scenario would not harm the system. In such situations, we could ask \texttt{MS Copilot} to prioritize the risks to perform a more accurate scenario selection. % that exhibit some kind of risks. 
%At last, we have to mention that we did not identify "no risks" for the scenarios as we believe that the risks already provide enough information for the selection of a given scenario.

\begin{table}[ht]
% \vspace{-0.3cm}
\caption{Risks and sensitivity points}\label{tab:risk-sensitivity}
\vspace{0.1cm}
\scalebox{0.75}{
\begin{tabular}{p{0.10\columnwidth} | p{0.20\columnwidth} | p{0.20\columnwidth} |p{0.30\columnwidth} |p{0.30\columnwidth}| p{0.15\columnwidth}|}
\toprule
Groups & Risks (students) & Risks (Copilot)  & Sensitivity points (students) & Sensitivity points (Copilot) & Scenarios selected with risks\\
%\tabularnewline
\hline
G1 & 1.1, 1.4, 3.1, 4.1 & 1.1, 1.4, 2.2, 2.3, 3.1, 3.3, 4.1, 4.4, 6.3 & Database, backup security, database scalability & Database, backup security, database scalability, Authentication, data cache, system workload, horizontal scalability, network security & 3.3 \\
\hline
G2 & 1.1, 1.3, 1.4, 2.1, 2.4, 3.1, 3.3, 3.4, 3.5 & 1.1, 1.2, 1.4, 1.5, 2.1, 2.3, 2.4, 3.1, 3.3, 3.4 & N/A & Data security, database performance, server concurrency, database replication, hot updates &1.2, 2.3, 3.3\\
\hline
G3 & 1.1, 1.2, 1.3, 1.4, 1.5, 2.1, 2.2, 2.3, 3.1, 3.2, 3.3, 3.4, 4.1, 4.2, 4.3 & 1.1, 1.2, 1.3, 1.4, 1.5, 2.1, 2.2, 2.3, 3.1, 3.2, 3.3, 3.4, 4.1, 4.2, 4.3 & Peak workload, latency, bottlenecks, database security, biometric security & Peak workload, latency, bottlenecks, database security, biometric security, network capacity, complexity of authentication methods, lack of infrastructure redundancy, integration between components & 1.1, 2.3, 3.2\\
\hline
G4 & 1.1, 1.6, 2.1, 2.4 & 1.1, 1.2, 1.6, 2.1, 2.4, 3.1 & Database & Database, encryption mechanism, load balancer, cache, 2FA multifactor authentication & 1.1, 1.6 \\
\hline
G6 & 1.1, 2.2, 2.4, 3.2 & 1.1, 2.2, 2.4, 3.2 & Database & Database & None \\
\hline
G7 & 1.1, 1.2, 3.1, 3.2 & 1.1, 1.2, 2.3, 3.1, 3.2 & Database encryption, database scalability & Database encryption, database scalability & 3.2 \\
\hline
G8 & 1.1, 1.2, 1.3, 2.2, 2.3, 3.1, 3.6 & 1.1, 1.2, 1.3, 2.2, 2.3, 3.1, 3.4, 3.5, 3.6 & Server availability, database caching, database encryption & Database caching, database encryption, authentication failures, server monitoring & 1.3, 2.2, 3.1, 3.4 \\
\bottomrule
\end{tabular}
}\vspace{-0.5cm}
\end{table}

\textbf{Prompt \#3. Tradeoff analysis:} In this step we prompted \texttt{MS Copilot} to perform a quality-attribute tradeoff analysis for the scenarios identified in the utility tree by each group. The results are shown in Table \ref{tab:tradeoff}. The first two columns indicate the qualities involved in each tradeoff identified by the students and by \texttt{MS Copilot}, while the third and fourth columns indicate the scenarios involved and selected during the tradeoff analysis by the students and \texttt{MS Copilot} as well. As it can be seen, \texttt{MS Copilot} identified more trade-offs between qualities in all cases. Therefore, the analysis performed by the LLM involves more scenarios for the tradeoff analysis. Also, in general terms, the scenarios chosen by \texttt{MS Copilot} are in line with the selection made by the students except in some cases, where MS Copilot considered that more scenarios could be selected (e.g. $G6$, $G7$). 

\begin{table}[ht]
% \vspace{-0.3cm}
\caption{Tradeoff analysis points}\label{tab:tradeoff}
\vspace{0.1cm}
\scalebox{0.75}{
\begin{tabular}{p{0.10\columnwidth} | p{0.35\columnwidth} | p{0.35\columnwidth} |p{0.25\columnwidth} |p{0.20\columnwidth}|}
\toprule
Groups & Tradeoff analysis (students) & Tradeoff analysis (Copilot)  & Scenarios affected by tradeoffs & Wining scenarios after tradeoff\\
%\tabularnewline
\hline
G1 & Performance-Security, Performance-Reliability & Performance-Security, Performance-Reliability, Security-Usability, Performance-Scalability, Security-Maintainability & Students: None  Copilot: 1.1, 1.2, 2.2, 2.3, 2.4, 4.1, 4.2, 5.2 & Students: 1.2, 2.1, 3.3, 4.2, 5.2, 6.6 Copilot: 1.2, 2.1, 3.3, 4.2, 5.2 \\
\hline
G2 & Security-Performance, Security-Availability & Security-Performance, Security-Availability, Performance-Availability & Students: 1.1, 1.3, 1.6  Copilot: 1.1, 1.3, 2.1, 2.3, 2.4, 2.6, 3.1, 3.6 & Students: 1.2, 1.6, 2.3, 2.6, 2.9, 3.3, 3.6 Copilot: 1.2, 1.6, 2.3, 2.6, 2.9, 3.6 \\
\hline
G3 & Scalability-Performance, Security-Performance, Availability-Performance, Authenticity-Compatibility & Scalability-Latency, Security-Performance, Availability-Complexity, Authenticity-Usability & Students: ALL  Copilot: ALL & Students: 1.1, 2.3, 3.2, 4.4 Copilot: 1.2, 2.3, 3.1, 4.3 \\
\hline
G4 & Security-Performance, Scalability-Performance & Security-Performance, Scalability-Performance, Scalability-Complexity, Performance-Consistency, Security-Usability, Availability-Complexity & Students: 1.1, 1.6, 2.2, 2.3, 3.1  Copilot: 1.2, 2.1, 2.4, 2.5, 3.1 & Students: 1.1, 1.6, 2.2, 2.5, 3.1 Copilot: 1.2, 2.1, 2.4, 3.1 \\
\hline
G6 & Performance-Interoperability, Security-Performance, Performance-Availability & Security-Performance, Interoperability-Complexity, Availability-Cost, Scalability-Simplicity & Students: ALL  Copilot: ALL & Students: 1.2, 2.1, 3.4  Copilot: 1.2, 2.1, 2.3, 3.4 \\
\hline
G7 & Security-Performance & Security-Performance, Performance-Scalability, Scalability-Cost, Security-Usability, Performance-Complexity, Scalability-Maintainability & Students: 1.1, 1.2, 1.3  Copilot: ALL & Students: 1.3  Copilot: 1,2, 1.3, 2.3, 3.1, 3.3 \\
\hline
G8 & Reliability-Availability, Reliability-Performance Efficiency, Security-Performance & Reliability-Performance Efficiency, Security-Performance Efficiency, Scalability-Maintainability, Availability-Cost, Security-Usability & Students: ALL  Copilot: 1.1, 1.2, 1.3, 2.1, 2.2., 2.3, 3.1, 3.2, 3.3., 3.4 & Students: 1.3, 2.1, 2.2, 2.4 Copilot: 1.2, 1.3, 2.1, 3.1 \\
\bottomrule
\end{tabular}
}\vspace{-0.5cm}
\end{table}

\textbf{Prompt \#4. Selection of scenarios:} In Table \ref{tab:atam} we show, for each group and quality attribute, the identifiers of the scenarios selected by the students and by \texttt{MS Copilot}. We marked in red those cases where the scenarios chosen by the students and by the LLM were very different, and in green those that exhibited a very good match. In the rest of the cases, there was some degree of scenario matching, but also some differences between the students and \texttt{MS Copilot}. We must remark that in the case of $G6$ the matching of the scenarios was almost perfect. Overall, only in $4$ cases out of $30$ the scenarios selected by \texttt{MS Copilot} diverged from the selection of the students. Thus, in general terms, we can consider that the use of an LLM was beneficial to evaluate and select scenarios. Finally, we asked \texttt{MS Copilot} to select the most suitable scenarios for a specific quality attribute and area of the system affected by it, but %if ask Copilot to select only one scenario, 
the LLM often selected only one scenario. This is an example of the %prompt engineering 
challenges that still exist %Therefore, this is an aspect that affect the prompt engineering challenges 
when asking an LLM to evaluate and select quality scenarios.

\begin{table}[ht]
% \vspace{-0.3cm}
\caption{Quality-attribute scenarios selected}\label{tab:atam}
\vspace{0.1cm}
\scalebox{0.75}{
\begin{tabular}{p{0.15\columnwidth} | p{0.18\columnwidth} | p{0.15\columnwidth} |p{0.15\columnwidth} |p{0.15\columnwidth} | p{0.15\columnwidth} |p{0.15\columnwidth} |p{0.15\columnwidth} |}
\toprule
Groups and scenarios & Quality 1 & Quality 2  & Quality 3 & Quality 4 & Quality 5 & Quality 6 & Quality 7\\
%\tabularnewline
\hline
G1 & Security & Security & Reliability & Performance Efficiency (data access) & Performance Efficiency (orders) & & \\
Sc. students & \textcolor{green}{1.2} & \textcolor{green}{1.2} & \textcolor{green}{3.3, 4.2} & \textcolor{green}{5.2} & \textcolor{red}{6.2} & &\\
Sc. Copilot & \textcolor{green}{1.2} & \textcolor{green}{2.1} & \textcolor{green}{3.3, 4.2} & \textcolor{green}{5.2} & \textcolor{red}{None} & &\\
\hline
G2 & Security (database) & Security (server) & Performance (database) & Performance (server) & Performance (orders) & Availability (database) & Availability (orders)\\
Sc. students & 1.2 & 1.6 & 2.3 & \textcolor{red}{2.6} & 2.9 & \textcolor{red}{3.3} & 3.6\\
Sc. Copilot & 1.2, 1.3 & 1.5, 1.6 & 2.2, 2.3 & \textcolor{red}{2.5} & 2.8, 2.9 & \textcolor{red}{3.1, 3.2} & 3.4, 3.6\\
\hline
G3 & Scalability & Security (database) & Security-Authenticity & Availability & &  & \\
Sc. students & 1.2 & 2.3 & 3.2 & \textcolor{red}{4.1, 4.2} & & &\\
Sc. Copilot & 1.1, 1.2 & 2.2, 2.3 & 3.1, 3.2 & \textcolor{red}{4.3, 4.4} & & &\\
\hline
G4 & Security (database) & Security (client) & Scalability (GW) & Scalability (database) & Performance (database)  & & \\
Sc. students & \textcolor{green}{1.1j} & \textcolor{red}{1.6} & \textcolor{green}{2.2} & \textcolor{green}{2.5} & \textcolor{green}{3.1} & &\\
Sc. Copilot & \textcolor{green}{1.1} & \textcolor{red}{1.4} & \textcolor{green}{2.2} & \textcolor{green}{2.5} & \textcolor{green}{3.1} & &\\
\hline
G6 & Interoperability & Security & Availability &  &   & &\\
Sc. students & \textcolor{green}{1.2} & \textcolor{green}{2.1} & \textcolor{green}{3.4} & & & &\\
Sc. Copilot & \textcolor{green}{1.2} & \textcolor{green}{2.1} & \textcolor{green}{3.4} & & & &\\
\hline
G7 & Security & Performance & Scalability &  &  & &\\
Sc. students & 1.3 & \textcolor{green}{2.2, 2.3} & 3.1 & & & &\\
Sc. Copilot & 1.2, 1.3 & \textcolor{green}{2.2, 2.3}  & 3.2, 3.3 & & & &\\
\hline
G8 & Reliability & Performance Efficiency & Security &  &  & &\\
Sc. students & \textcolor{green}{1.3} & 2.1, 2.2 & 3.1, 3.2, 3.4 & & & &\\
Sc. Copilot & \textcolor{green}{1.3} & 2.1  & 3.2, 3.3 & & & &\\

%\tabularnewline

\bottomrule
\end{tabular}
}\vspace{-0.5cm}
\end{table}

%In addition, the scenarios selected by the students were based on an analysis of possible risks for the scenarios, identification of sensitivity points in the architecture that could affect the scenarios, and a tradeoff analysis between quality attributes. % in some of the scenarios. 
%Thus, we used three additional prompts to query the LLM about these three aspects and validate the scenarios suggested by \texttt{MS Copilot }in Table \ref{tab:atam}.

In the case of the risks and sensitivity points identified by the students and \texttt{MS Copilot} in Table \ref{tab:risk-sensitivity}, \texttt{MS Copilot} was able to identify additional risks and sensitivity points in most cases affecting more scenarios than those of the students. However, not all the new risks identified by \texttt{MS Copilot} had the same degree of severity. Some risks are important, while others can be overlooked in the selection of a given scenario. The last column of the table indicates scenarios selected by the students but affected by risks identified by the \texttt{MS Copilot}. Due to space constraints, we cannot provide a detailed classification of the severity of the risks for each scenario. Therefore, a thorough analysis of the severity of each risk must be made in a further study. In addition, \texttt{MS Copilot} identified new sensitivity points in the architecture that the students did not raise. This is particularly important for identifying potential quality tradeoffs, hidden risks, or extending the students' analysis by including critical system design areas. 

With regard to the tradeoff analysis results shown in Table \ref{tab:tradeoff}, \texttt{MS Copilot} identified in all cases more tradeoffs than the students. Nevertheless, some of these tradeoffs belong to certain qualities (e.g., cost, usability) that the students were told to ignore due to the lack of cost numbers or the difficulty to evaluate certain qualities (e.g., usability). Although \texttt{MS Copilot} was able to suggest more scenarios affected by these tradeoffs, a deeper analysis would be recommended to assess the accuracy of the tradeoffs. Finally, from the scenarios selected from the tradeoff analysis, our initial results reveal that \texttt{MS Copilot} matched the students' results in several cases, while in others suggested different scenarios. For example, in group $G6$, the students reported only one scenario ($1.3$) for a tradeoff between two qualities while \texttt{MS Copilot} indicated $5$ scenarios from tradeoffs between $6$ pairs of qualities. The results of the students are independent of the final numbers of scenarios selected. In other cases, such as group $G3$, all scenarios except one are different in the tradeoff analysis between the students and \texttt{MS Copilot}, probably because the LLM identified $6$ tradeoffs versus the two tradeoffs reported by the students. 

\vspace{-0.2cm}\section{Conclusions and Future Work}\label{sec:conclusions}\vspace{-0.1cm}
In this paper, we have investigated how LLMs can provide support and improve the results of an ATAM architecture evaluation process carried out by students of a software architecture course. Overall, \texttt{MS Copilot} was able to produce more information for key phases of ATAM. More specifically, the LLM produced more results in the identification of risks, sensitivity points and tradeoff analysis, which contributes to improve the outputs of the evaluation process.

The results so far confirm our hypothesis about the usefulness of LLMs in the assessment of ATAM scenarios, identifying additional information to support the architecture evaluation of quality requirements. In future work, we plan to explore two paths of work: (i) using LLMs to create and rank quality scenarios, (ii) performing a deeper analysis of the role of concrete risks, sensitivity points, and accuracy of tradeoffs. Finally, alternative LLMs can be explored to cross-validate the results of \texttt{MS Copilot } and to see whether contextual knowledge from previous evaluations can affect the reported results.

%\vspace{-0.1cm}
\subsubsection{Data Availability} \label{sec:data-availability}\vspace{-0.3cm}
We provide an anonymized site with the data used in the paper: \url{https://anonymous.4open.science/r/archevaluation-llms-1FDE/}.  

\vspace{-0.1cm}
\small{
\bibliography{references}
}

\end{document}